\documentclass[twocolumn]{aastex631}

\usepackage{xcolor}

\usepackage{amsmath}
\graphicspath{{./}{figures/}}

\shorttitle{Millimeter Emission from Interacting Supernovae}
\shortauthors{Yadlapalli et al.}

\begin{document}

\title{Models of Millimeter and Radio Emission from Interacting Supernovae}
\author[0000-0003-3255-4617]{Nitika Yadlapalli}
\author[0000-0002-7252-5485]{Vikram Ravi}
\affiliation{Cahill Center for Astronomy and Astrophysics, MC 249-17 California Institute of Technology, Pasadena CA 91125, USA}

\author[0000-0002-9017-3567]{Anna Y. Q.~Ho}
\affiliation{Miller Institute for Basic Research in Science, 468 Donner Lab, Berkeley, CA 94720, USA}
\affiliation{Department of Astronomy, University of California, Berkeley, Berkeley, CA, 94720, USA}
\affiliation{Lawrence Berkeley National Laboratory, 1 Cyclotron Road, MS 50B-4206, Berkeley, CA 94720, USA}

\begin{abstract}

   This work utilizes established models of synchrotron-powered light curves for core-collapse supernovae in dense circumstellar environments, namely type IIn and Ibn, to demonstrate the potential for detecting millimeter emission from these events. The progenitor types of these supernovae are still an open question, but using the synchrotron light curves as probes for the circumstellar environments could shed light on the mass-loss histories of the progenitors and discern between different theories. Observations in millimeter bands are particularly fruitful, as they probe regions at smaller radii and higher ambient densities, where centimeter emission tends to be self-absorbed. In our application of these light curves, we explore a diversity of progenitor types and mass-loss profiles to understand their effects on the light curve shapes. Additionally, we fit model parameters to the 8\,GHz light curve of type IIn supernova 2006jd and then create millimeter light curves using these parameters to show the possibility of detecting an early millimeter peak from such an event. We predict that next generation millimeter surveys will possess the capability to detect nearby and extreme events. However, there is a pressing need for millimeter follow-up of optically discovered interacting supernovae to more completely sample the true population.
    
\end{abstract}

\section{Introduction}

\subsection{Millimeter Transients}

To date, the transient millimeter sky remains largely unexplored. In recent years though, millimeter transients have been discovered on timescales from minutes to months. The first blind survey for millimeter transients was conducted by \cite{Whitehorn2016} using the South Pole Telescope (SPT), which surveyed 100\,deg$^2$ of sky and discovered only one candidate transient at $<3\sigma$ significance. An improved transient search was conducted with the SPT-3G camera, covering a 1500\,deg$^2$ with deeper sensitivity \citep{Guns2021}. This search yielded 10 unique transient sources of which 8 were identified as flaring stars and 2 were identified as extragalactic events of unknown origin. Other blind detections of millimeter transients include 3 candidate stellar flares found serendipitously by the Atacama Cosmology Telescope \citep[ACT;][]{Naess2020}. Though these works imply a higher event rate for millimeter stellar flares than previously known, the rates of extragalactic millimeter transients is still an open question.

Extragalactic transient millimeter emission arises from energetic synchrotron sources. \cite{Metzger2015} and more recently \cite{Eftekhari2021} predict that wide-field millimeter surveys would be most sensitive to reverse shock emission from gamma-ray bursts (GRBs), such as those detected by \cite{DeUgartePostigo2012} and \cite{Laskar2013}. They also predict millimeter detections of tidal disruption events (TDEs), such as on-axis jetted sources like Swift J164449.3+573451 \citep{Zauderer2011} or off-axis jetted sources like IGR J12580+0134 \citep{Yuan2016}. Optimistically, they additionally hope for a few detections of fast blue optical transients (FBOTs), such as nearby, luminous, millimeter transient AT2018cow \citep{Ho2019_cow}. However, detections of typical core-collapse supernovae (SNe) at millimeter wavelengths are scarce. Here we demonstrate the potential of early targeted observations of interacting SNe to dramatically increase the number of detections and probe the environments in which these explosions occur.

\subsection{Millimeter Detectable Supernovae}

Core-collapse supernovae (CCSNe) result from the deaths of massive stars. There exists a broad diversity of flavors of CCSNe, usually arising from different progenitor types \citep[see][and references within]{Smartt2009}. The most commonly observed supernovae are type II-P, characterized by a plateau in the light curve following the initial decline from peak brightness. The plateaued light curve is powered by expansion of the photosphere due to recombination in the hydrogen envelope of a red supergiant (RSG) progenitor. Type II-L supernovae, characterized by a linearly declining light curve rather than a plateaued one, are similar to type II-P but arise from RSG progenitors that have stripped hydrogen envelopes. Type Ibc/IIb supernovae are also the products of stripped envelopes, but arise from massive Wolf-Rayet (WR) stars rather than RSGs. Interacting supernovae, however, are unique because they have no distinctive progenitor type.

\begin{figure}[t]
    \centering
    \includegraphics[width=\columnwidth]{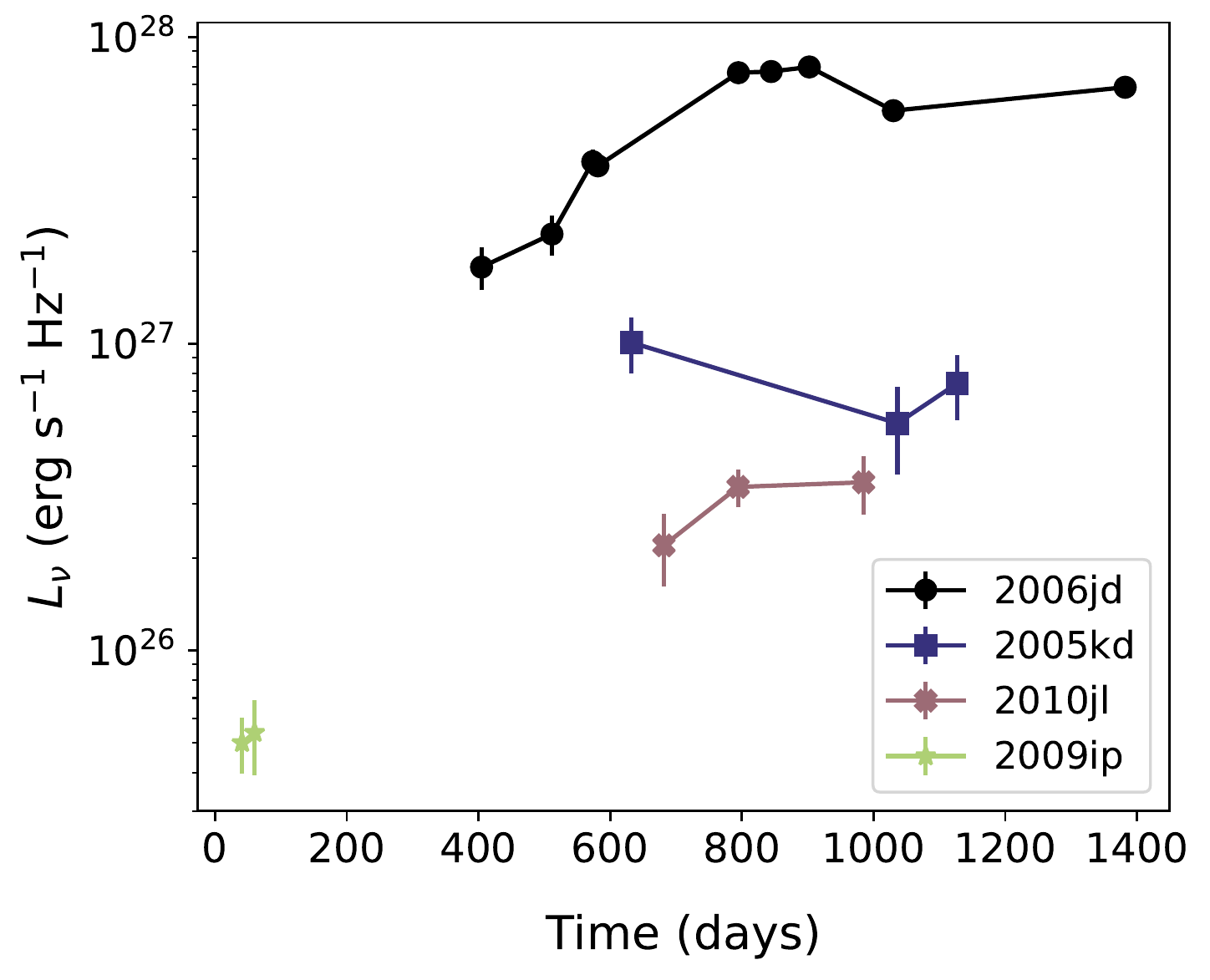}
    \caption{Reproduction of 8.5 GHz light curves of four type IIn SNe. This is meant to show the diversity of light curves that have been observed, highlighting the challenge in constraining a single progenitor model for these events. The data for these was taken from \cite{Chandra2012}, \cite{Dwarkadas2016}, \cite{Chandra2015}, and \cite{Margutti2014}.}
    \label{fig:IIn_8GHz}
\end{figure}

Interacting supernovae are identified by their bright, narrow spectral lines that arise from shock interaction with a very dense circumstellar medium \citep[CSM;][]{Chugai1990}. Type IIn SNe spectra are dominated by hydrogen lines whereas type Ibn SNe show weak hydrogen but strong helium lines \citep[see][and references within for more detail]{Smith2017}. High velocity shocks, reaching up to a few percent of $c$, propagating through regions of density orders of magnitude greater than ISM ($n_e \sim 10^6$\,cm$^{-3}$)  are optimal conditions for producing bright radio and millimeter emission. Millimeter emission is especially important, as it peaks at early times when the shock radius is smaller and the CSM density is higher. Early time radio emission has been detected from type Ib/IIb SNe, such as the nearby SN 1993J \citep{Pooley1993}, or from type Ibc events, such as SN1998bw \citep{Kulkarni1998} or SN2009bb \citep{Soderberg2010}. A handful of Ib/IIb SNe have been detected in millimeter observations. The very nearby events SNe iPTF13bvn \citep{Cao2013} and SN2011dh \citep{Horesh2013}, for example, were detected with the Combined Array for Research in Millimeter-wave Astronomy (CARMA), and were essential to constraining the early evolution of the shock radius and providing estimates of the pre-supernova mass loss rate. To date, only a handful of type IIn supernovae have been detected at radio wavelengths. Figure 7 in \cite{Chandra2018} summarizes detections of IIn supernovae at 8 GHz -- we reproduce a few of these light curves here for reference in Figure~\ref{fig:IIn_8GHz}. Notably, nearly all of the detections are made over a hundred days post-explosion. \cite{Bietenholz2021} show that the luminosity-risetime parameterization of radio emission from type IIn SNe is characterized by a significantly later time to peak than that of type Ib/c or other type II SNe. Thus observations at higher frequencies would prove useful in detecting the peak of the light curve at an earlier time. Type IIn and Ibn SNe have yet to be detected in the millimeter, but are likely to  be millimeter-bright owing to their especially high CSM densities. 

\subsection{Diversity of Progenitor Models for Interacting Supernovae}

The nature of the progenitors for interacting supernovae is still unknown, owing to the spectacularly high observed mass-loss rates of up to 0.1\,$M_{\odot}$/yr and little understanding of the mechanisms driving this. Luminous blue variables (LBV), among the most luminous known stars, are commonly theorized to be IIn progenitors. \cite{GalYam2007} provides the most direct evidence for this theory with pre-explosion \textit{Hubble Space Telescope} observations of IIn SN 2005gl that revealed a source coincident with the supernova position with a high luminosity that could only be explained by an LBV. Detections of pre-supernova outbursts also hint towards an LBV origin for IIn SN. Famously, SN 2009ip, originally thought to be a type II SN, was discovered to be a bright LBV-like outburst prior to a true IIn supernova explosion in 2012 \citep{Mauerhan2013}. The evolutionary pathway of LBVs, however, is still debated, casting doubt on their candidacy as IIn progenitors. \cite{Humphreys1994} suggest that LBVs are O-type stars experiencing a period of high mass loss en route to becoming Wolf-Rayet (WR) stars; stars in this evolutionary stage would not be expected to explode as supernovae. \cite{Smith2015} offers a model of LBVs arising from binary evolution, which better fits a picture of LBVs as IIn progenitors. Blue supergiants (BSGs) have also been hypothesized to be IIn progenitors as models of BSG explosions are consistent with the types of precursor outbursts observed in events such as SN 2009ip and SN 2010mc \citep{Smith2014}. Additionally, spatially resolved observations of the gas around VY CMa indicates that red supergiants (RSGs) may also have CSM structures that could give rise to IIn supernovae \citep{Smith2009}. Thus, the ability to probe the CSM structure around observed supernovae has the power to shed light on the nature of their progenitors.

Observations of synchrotron powered light curves around the time of their peaks enable measurements of the radii and electron densities of the CSM the supernovae is interacting with at that time, using equations published in works such as \cite{Duran2013} and \cite{ho2021}. Constraining the shock velocity and the CSM density by these measurements allows constraints to be placed on the properties and mass-loss history of the progenitor, narrowing down viable progenitor models. As spherically expanding synchrotron sources will peak earlier in the millimeter than at lower frequencies, early millimeter observations are especially important in measuring the CSM density close to the surface of the progenitor, giving us a glimpse into the behavior of the progenitor in its final days to months.

In this paper, we model light curves of radio and millimeter emission from interacting supernovae. Section~\ref{sec:model_summary} goes through a summary of the emission and absorption models we use. Section~\ref{sec:typeII} presents applications of this model to generate synthetic light curves of type IIn supernovae and Section~\ref{sec:rate} shows the potential for detecting these events with next generation cosmic microwave background (CMB) surveys. We conclude in Section~\ref{sec:conclusion}.

\section{Model Summary}\label{sec:model_summary}

Radio emission from supernovae originates in synchrotron radiation from free electrons in the CSM that are accelerated to relativistic energies by the forward supernova shock. The brightness and evolution of the emission is dependent on the density and structure of the CSM, as will be seen in the equations presented below as well as in the light curves presented in the following section. The aim of this section is not to present novel emission models, but to give an overview of well-established models that we utilize. We reproduce many relevant equations from the works cited in this section to highlight specific model parameters that we explore in the rest of this work.

For an interacting core-collapse supernova, the progenitor star undergoes pre-supernova mass loss in the decades leading up to the explosion \citep[for example]{Smith2017}. The resulting CSM density can be modelled as power law described by $\dot{M}$, the mass loss rate, and $v_w$, the wind velocity, as shown in Equation~\ref{eq:rho_CSM}. $R_{*}$ is the radius of the progenitor star and $s$ describes the steepness of the CSM density profile. A steady wind can be modelled with $s=2$, but non-steady mass loss can lead to deviations from this value.Equations~\ref{eq:rho_CSM} and \ref{eq:rho_ej} and much of the following formalism is introduced and explained in detail in \cite{Chevalier1982}.

\begin{equation}
\rho_{CSM} = \frac{\dot{M}}{4\pi v_w  R_{*}^{2}} \left( \frac{R_{*}}{R} \right) ^s
\label{eq:rho_CSM}
\end{equation}

Following the collapse of the core, a radiation dominated shock travels through the progenitor star until the shock reaches a radius that is optically thin to high energy photons. When the shock breaks out of the stellar surface, these high energy photons propagate through the CSM, giving rise to an ionization front. Works such as \cite{Nakar2010} and \cite{Kochanek2019} go into detail about models and observations of the shock breakout, but we consider a simplified case for this work.

The expansion of the ionization front is bounded by the speed of light and is given by Equation~\ref{eq:Rbo} where $t$ is the elapsed time, $Q$ is the number of ionizing photons per second, and $m_p$ is the proton mass. The maximum radius of the ionization front, given in equation \ref{eq:Rbo_max}, is dictated by $T$, the duration of the breakout pulse. Assuming the shock breaks out at the stellar surface, the duration of the breakout pulse can be approximated by the light crossing time of the progenitor, $T = R_*/c$. For extreme CSM densities, the shock breakout may occur in the wind, but we ignore this possibility for now. We additionally ignore any recombination in the CSM.

\begin{equation}
    R_{ion} = \frac{Q m_p v_w}{\dot{M}}t \leq ct
    \label{eq:Rbo}
\end{equation}

\begin{equation}
    R_{ion, max} = \frac{Q m_p v_w}{\dot{M}} T
    \label{eq:Rbo_max}
\end{equation}

After the shock breakout, a forward shock continues to expand through the CSM. As the fastest moving unshocked ejecta reaches the decelerating shock front, a second shock known as the reverse shock forms. In the reference frame of the forward shock, the reverse shock propagates backwards and reheats the ejecta. It should be noted though that in the observer frame, the reverse shock is mostly moving radially outwards. The distinction between these two reference frames is important.

The boundary between the reverse shock and the forward shock is a contact discontinuity. We follow the discussion in \cite{Chevalier1994} and Section 5.6 of \cite{Vink2020} for their derivations for shock radius and velocity. The shock behavior can be modelled by approximating the forward and reverse shocks as a thin shell, meaning we assume that the radius of both the forward and reverse shock is approximately equal to the radius of the contact discontinuity. The radius and velocity of this shell is calculated by balancing the pressure difference across the contact discontinuity with the amount of deceleration. The reverse shock pressure near the contact discontinuity is dependent on the density profile of the supernova ejecta near the surface, $\rho_{ej}$, where $n$ defines the power-law index of the density profile.

\begin{equation}
\rho_{ej} = \rho_o \left( \frac{t}{t_o} \right)^{-3} \left( \frac{v_{s,o}t}{r} \right)^n
\label{eq:rho_ej}
\end{equation}

The inner regions of the ejecta are assumed to have a flat density distribution. The initial density and velocity of this core are described by constants $\rho_{o} t_{o}^3$ and $v_{s,o}$, which are derived using total ejecta mass, $M_{ej}$, and the kinetic energy of the ejecta, $E_{kin}$.

\begin{equation}
v_{s,o} = \sqrt{\frac{10}{3} \frac{(n-5)}{(n-3)} \frac{E_{kin}}{M_{ej}}}
\label{eq:vso}
\end{equation}

\begin{equation}
\rho_{o} t_{o}^3 = \frac{M_{ej}^{5/2} E_{kin}^{-3/2}}{\frac{4\pi}{3} \left( \frac{n}{n-3} \right) \left( \frac{10}{3} \frac{n-5}{n-3} \right)^{3/2}}
\label{eq:pt3}
\end{equation}

The radius and velocity of the shell are then derived to be equations~\ref{eq:Rs} and \ref{eq:Vs}, respectively.

\begin{equation}
    \begin{aligned}
    R_s(t)  =
    \left[ \frac{4\pi (3-s) (4-s) \rho_o t_o^3 v_{s,o}^n v_w R_{*}^{2-s}}{(n-4)(n-3)\dot{M}} \right]^{1/(n-s)} \\ \times \ t^{(n-3)/(n-s)}
    \end{aligned}
    \label{eq:Rs}
\end{equation}

\begin{equation}
    V_s(t)  = \frac{dR_s(t)}{dt} \propto t^{(3-s)/(2-n)}
    \label{eq:Vs}
\end{equation}

As the shock expands through the ionized CSM, the free electrons are Fermi-accelerated through the shock to relativistic speeds with a power law distribution of energies as described by equation~\ref{eq:NE}. The value of $p$ is dependent on details of the shock acceleration process. Table 1 in \cite{Chevalier1998} compiles a list of SNe with observed indices. Many of these SNe have power law indices close to 3, though some exhibit smaller values ranging as low as 2. In this work, we will consider the case of $p=3$, but the results presented here do not change significantly based on the value of $p$.

\begin{equation}
    N(E)\propto E^{-p}
    \label{eq:NE}
\end{equation}

We assume that all electrons are shocked to relativistic speeds and that the electrons always contribute a constant fraction, $\epsilon_e$, of the total energy density of the shocked gas. Applying the Rankine-Hugoniot jump conditions for the case of a fast shock \citep[see][for example]{Draine}, the post-shock energy density is 

\begin{equation}
    u_{tot} = \frac{9}{8}\rho_{\text{CSM}} V_s^2
    \label{eq:utot}
\end{equation}

The strength of the emission also depends on the magnetic field behind the shock. We assume that the post-shock magnetic field energy density is a constant fraction, $\epsilon_B$, of the total energy density. 

\begin{equation}
    B = \sqrt{8 \pi u_{tot} \epsilon_B}
    \label{eq:B}
\end{equation}

For synchrotron sources, the frequency at which the source has an optical depth of unity is known as the synchrotron self-absorption (SSA) frequency, $\nu_a$. Above this frequency, the source will be optically thin and below this frequency the source will be optically thick. For a source of a given age, we can also define the cooling frequency, $\nu_c$. This is defined as the characteristic frequency of an electron that has radiated an amount of energy equivalent to its total energy. It is defined in equation \ref{eq:nu_c}, where $m_e$ is the electron mass, $q_e$ is the elementary charge, and $\sigma_T$ is the Thomson cross section.

\begin{equation}
    \nu_c = \frac{18 \pi m_e c q_e}{\sigma_T^2 B^3 t^2}
    \label{eq:nu_c}
\end{equation}

The shape of the source spectrum depends on the ordering of $\nu_a$ and $\nu_c$. The supernova shock will begin in the fast cooling regime, where $\nu_c < \nu_a$. Referencing equations in Appendix C of \cite{ho2021}, we can write expressions for $\nu_a$ and the corresponding peak luminosity, $L_a$, for the fast cooling case.

\begin{equation}
    \nu_a = R_s^{1/4} B^{3/4} t^{-1/4} \xi^{-1/8} \eta^{7/8}
\end{equation}

\begin{equation}
    L_a = 4 \pi R_s^{21/8} B^{11/8} t^{-5/8} \xi^{-5/16} \eta^{35/16} \zeta
\end{equation}

As the shock expands, $\nu_a$ decreases with time while $\nu_c$ increases. Eventually, the shock will enter the slow cooling regime where $\nu_c > \nu_a$. In this regime, $\nu_a$ and $L_a$ evolve as below.

\begin{equation}
    \nu_a = R_s^{2/7} B^{9/7} \eta
\end{equation}

\begin{equation}
    L_a = 4 \pi R_s^{19/7} B^{19/7} \eta^{5/2} \zeta
\end{equation}

The constants $\xi$, $\eta$, and $\zeta$ are defined as follows.

\begin{equation}
    \xi = \frac{\sigma_T^2}{18 \pi m_e c q_e}
\end{equation}

\begin{equation}
    \eta = \left( \frac{\sigma_T}{12 \pi^2 m_e^2 c}\right)^{2/7} \left( \frac{q_e}{2 \pi m_e c}\right)^{1/7} 
\end{equation}

\begin{equation}
    \zeta = \frac{1}{3} \left( 2 \pi m_e \right)^{3/2} \left( \frac{c}{q_e}\right)^{1/2}
\end{equation}

In both the slow and fast cooling regime, the optically thick part of the spectrum goes as $L_\nu \propto \nu^{5/2}$. The optically thin part of spectrum goes as $L_\nu \propto \nu^{-p/2}$ in the case of fast cooling.  When the shock transitions to slow cooling, the optically thin part of the spectrum goes as $L_\nu \propto \nu^{-(p-1)/2}$ for $\nu_a < \nu < \nu_c$ and shifts to $L_\nu \propto \nu^{-p/2}$ for $\nu > \nu_c$. See Spectra 2 and 3 within Figure 1 of \cite{Granot2002} for a visualization of this. The functional definition of these spectra is given in equation C17 of \cite{ho2021}.

The emission is additionally absorbed in the ionized CSM due to free-free absorption (FFA). A frequency dependent correction must by applied to the spectrum by multiplying by $e^{-\tau_{\nu}^{\text{FFA}}}$, where $\tau_{\nu}^{\text{FFA}}$ is given in equation \ref{eq:tau_FFA}. In the calculation of the free-free opacity, the value of $\kappa_{\nu}^{\text{FFA}}$ is derived in \cite{Panagia1975}.

\begin{equation}
    \tau_{\nu}^{\text{FFA}} = \int_{R_s}^{R_{ion}} \kappa_{\nu}^{\text{FFA}} n_e n_i ds
    \label{eq:tau_FFA}
\end{equation}

\begin{equation}
    \kappa_{\nu}^{\text{FFA}} = 4.74 \times 10^{-27} \left( \frac{\nu}{\text{1 GHz}} \right)^{-2.1} \left( \frac{T_e}{10^5 \text{K}} \right)^{-1.35}
    \label{k_FFA}
\end{equation}

\section{Application to Type II\lowercase{n}/I\lowercase{bn} Supernovae} \label{sec:typeII}

\subsection{Model Light Curves}

\begin{figure*}[htp]
    \centering
    \includegraphics[width=0.95\textwidth]{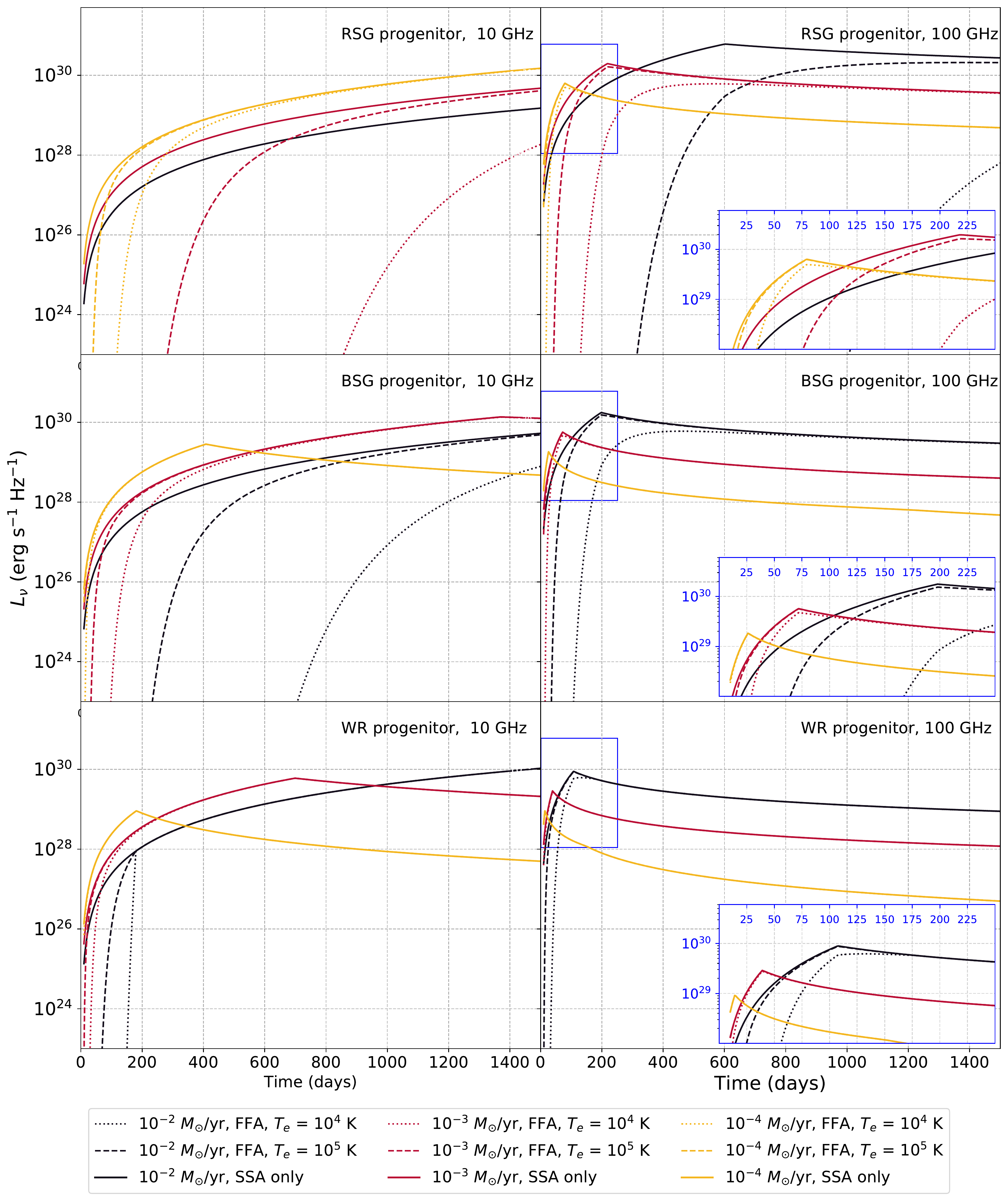}
    \caption{Light curves at 10 GHz and 100 GHz of synchrotron emission originating in the forward shock of an interacting supernova. Three absorption cases are shown here: SSA only, SSA + FFA assuming a $10^5$\,K CSM, and SSA + FFA assuming a $10^4$\,K CSM. The extent of the ionized region contributing to FFA is calculated given the shock breakout pulse properties. We use progenitor radii of  500\,$R_{\sun}$ (RSG), 70\,$R_{\sun}$ (BSG),and 5\,$R_{\sun}$ (WR) and pulse luminosities as shown in Figure 5 of \cite{Nakar2010}. We use steady wind velocities of 20\,km/s (RSG), 250\,km/s (BSG), and 1000\,km/s (WR). The RSG and BSG scenarios represent type IIn SNe while the WR scenario represents a type Ibn SNe. For the 100\,GHz light curves, we show a zoomed in panel of the first 250\, days to better visualize the time scale of the millimeter peaks. For scale, $\sim 10^{28}$\,erg\,s$^{-1}$\,Hz$^{-1}$ is about 1\,mJy at $\sim$100\,Mpc.}
    \label{fig:IIn/Ibn}
\end{figure*}

In this section, we present radio light curves of forward shock emission arising from a variety of interacting supernovae conditions. Given the broad range of viable progenitor models for interacting supernovae, we consider three progenitor types: RSGs, BSGs, and WR stars. We expect the RSG and BSG progenitors to represent light curves for type IIn supernovae and WR progenitors to represent light curves for type Ibn supernovae. The synchrotron emission model we detail in Section 2 is applicable to both scenarios. Each progenitor model varies in its breakout radius, wind velocity, and shock breakout pulse luminosity. We adopt the same radii and breakout luminosities for the progenitor models as those used in \cite{Nakar2010}. We chose a characteristic RSG wind velocity of 20\,km/s based on data on nearby RSGs \citep{Mauron2011}. OB type stars have wind velocities on the order of their escape velocities, so for our BSG and WR model we choose wind velocities of 250 and 1000\,km/s respectively. We assume an explosion energy of $E=10^{51}$\,ergs and an ejecta mass of $M_{ej}=5$\,$M_{\odot}$. The fraction of energy in electrons and magnetic fields is fixed to $\epsilon_e = \epsilon_b = 1/3$.

Our choices in parameter exploration are in part informed by modelling work done by \citet{Moriya2014} to fit bolometric light curves of type IIn SNe. Figure 9 in this work shows a range of estimated mass-loss rates from various type IIn SNe -- from this, we choose to explore three different mass loss rates in this work $\dot{M} = 10^{-2}, 10^{-3},$ and $10^{-4}$\,$M_{\odot}$/year. \citet{Moriya2014} also infers the CSM density slope, $s$, for each of these light curves. We explore parameters between $1.6 < s < 2.2$, but refer to \citet{Moriya2012} for a more detailed discussion on the important effect of the density profile on the resultant optical light curves. We note too that we assume an infinite wind and a fixed CSM density slope. For piecewise varying CSM models, we refer readers to works such as \citet{Matsuoka2019}, which present hydrodynamical simulations of radio light curves resulting from shock interaction with a confined CSM.

Figure~\ref{fig:IIn/Ibn} shows light curves for combinations of the three progenitors and three mass-loss rates at two frequencies, 10\,GHz and 100\,GHz, and two ionized CSM temperatures, $T_{e} = 10^4$ and $10^5$\,K. Ejecta and CSM density profile indices are fixed to $n=10$ and $s=2$ in this example. The dotted and dashed lines in this figure represent emission that one may observe, taking into account free-free absorption. Models including only synchrotron self-absorption, shown by solid lines, are included for comparison. These light curves illustrate the limits on searching for lower frequency radio emission from interacting supernovae. For example, for a supernova exploding from an RSG progenitor with a mass-loss rate of $10^{-4}$\,$M_{\odot}$/year, akin to a star like VY CMa, the 10\,GHz emission could not reasonably be detected until around 200\,days for the coolest CSM temperature of $10^4$\,K. The 100\,GHz emission from the same event, however, peaks sooner than 100\,days, yielding information about the CSM structure much closer to the progenitor surface than the 10\,GHz emission. Similarly, in the BSG case, 100\,GHz observations are essential to observe any early time emission from progenitors with mass-loss rate with $10^{-2}$\,$M_{\odot}$/year. For WR progenitors, the high wind velocities and small progenitor radii decrease the CSM densities as well as limit the extent of the ionization front, making free-free absorption much less of an issue in observing type Ibn supernovae, compared to IIn. For all progenitor types however, the millimeter light curve peaks much earlier than the radio light curve. This is because as the source radius increases and the shock decelerates, the SSA frequency decreases. As we will show in the next section, applying a model fitting approach to gleaning information from the light curve provides the most discerning information when detections are made around the peak. Thus, even independent of the problem of absorption, studies of interacting supernovae would greatly benefit from early time millimeter observations.

\begin{figure}
    \centering
    \includegraphics[width=\columnwidth]{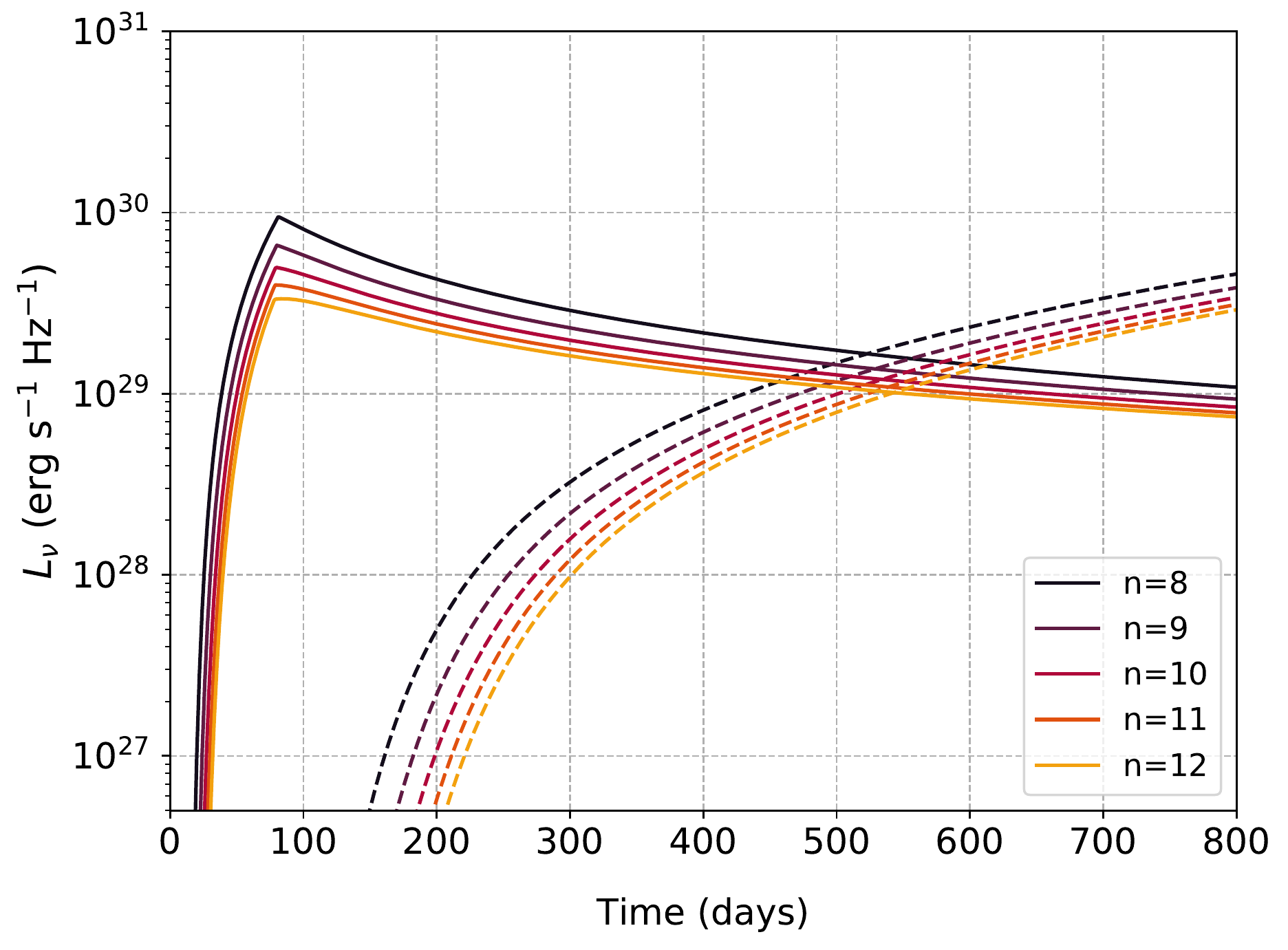}
    \includegraphics[width=\columnwidth]{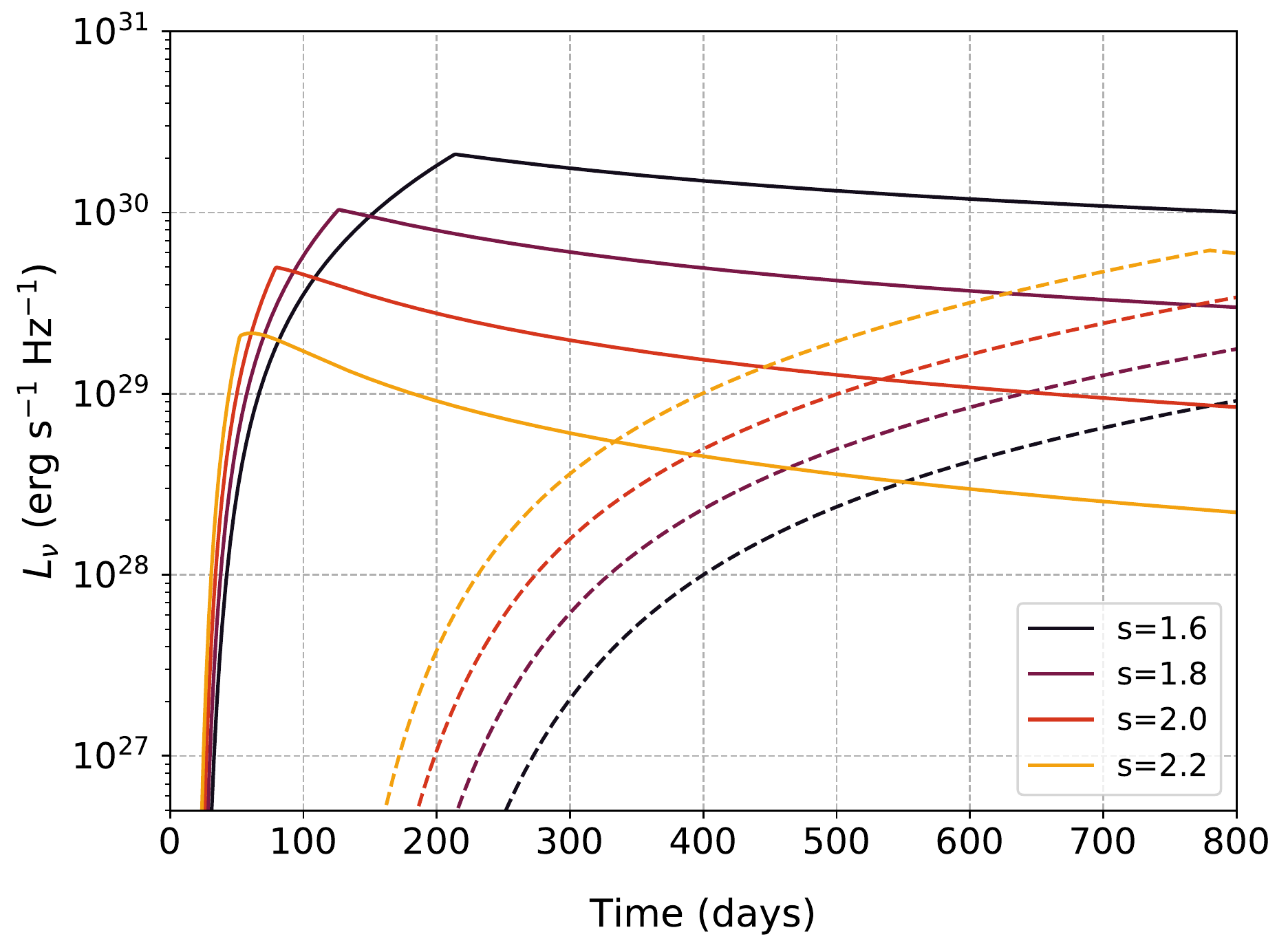}
    \includegraphics[width=\columnwidth]{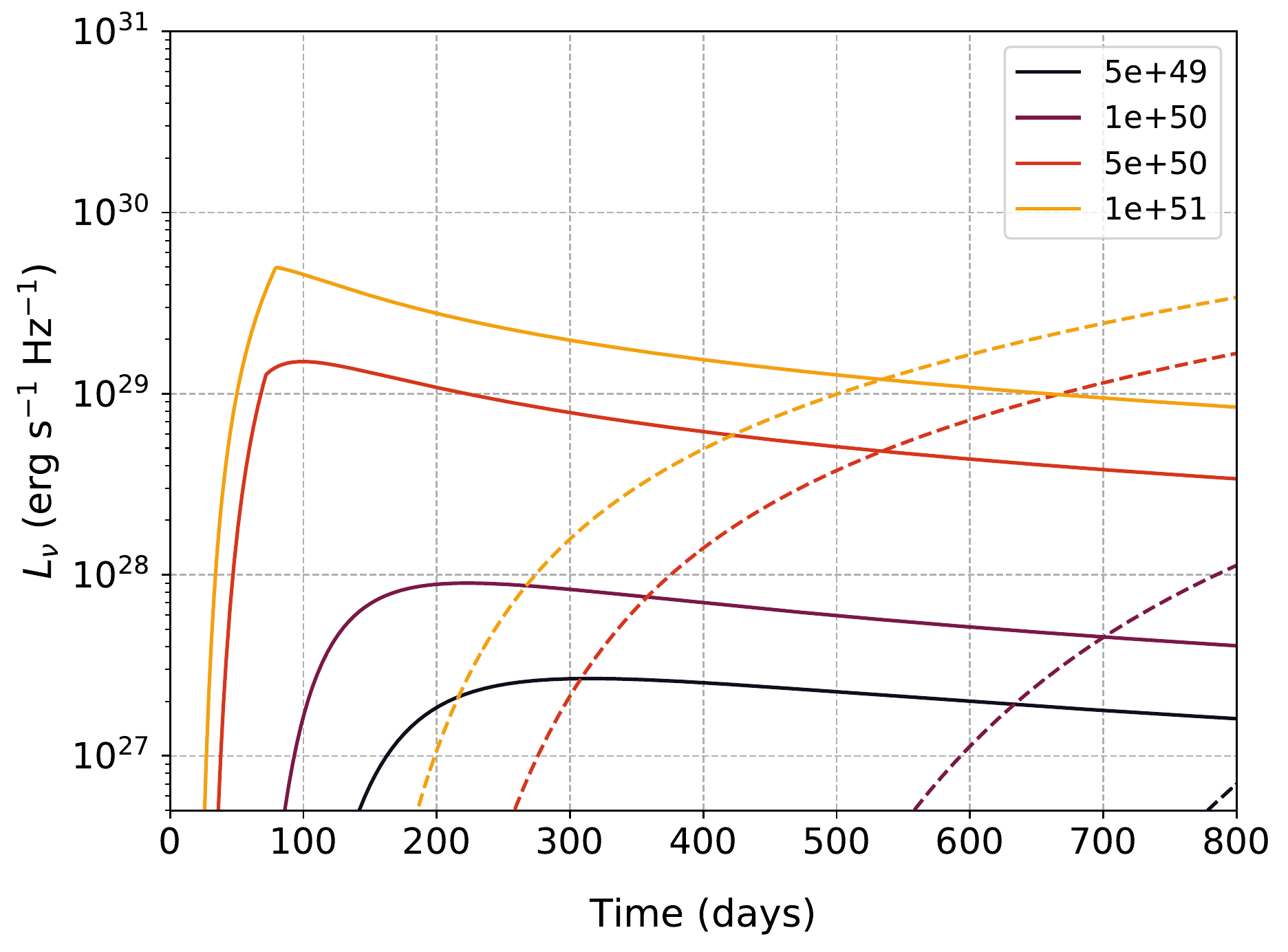}
    \caption{Light curves at 100\,GHz (solid lines) and 10\,GHz (dashed lines) showing the effects of different ejecta and CSM radial profiles. Adopting an RSG progenitor model and keeping constant $\dot{M}=10^{-3}$\,$M_{\odot}$/yr, $T_e = 10^4$\,K, $E=10^{51}$\,ergs, and $M_{ej}=5$\,$M_{\odot}$, the upper panel assumes $s=2$ and varies the ejecta profile index, $n$, while the middle panel assumes $n=10$ and varies the CSM profile index, $s$. The bottom panel assumes $s=2$ and $n=10$ and varies the explosion energy.}
    \label{fig:varyns}
\end{figure}

We also explore the effect of different ejecta and CSM density profiles as well as the initial explosion energy on the 100\,GHz and 10\,GHz light curves. To do this, we start with an RSG model and fix $\dot{M}=10^{-3}$\,$M_{\odot}$/yr, $T_e = 10^4$\,K, and $M_{ej}=5$\,$M_{\odot}$. In the upper panel of Figure~\ref{fig:varyns}, we fix $s=2$ and $E=10^{51}$\,ergs and vary the ejecta density power-law index in integer steps from 8 to 12. In the middle panel, we fix $n=10$ and $E=10^{51}$\,ergs, and vary the CSM density power-law index $s$ from 1.6 to 2.2 in steps of 0.2.  In the bottom panel, we fix $s=2$ and $n=10$ and vary the initial explosion energy between $5 \times 10^{49}$\,ergs $< E < 10^{51}$\,ergs. The trends discussed below hold for both the BSG and WR models as well, varying only in the exact values of the peak time and luminosity. For brevity, we show only plots produced for the RSG model.

Figure~\ref{fig:varyns} shows that the millimeter and radio light curves are most sensitive to changes in explosion energy and CSM density profile. The middle panel of Figure~\ref{fig:varyns} shows that for a fixed mass-loss rate, decreasing $s$ increases the peak luminosity of the millimeter light curve. This is because while a steady wind is modelled with $s=2$, a value of $s < 2$ represents that in the past, the progenitor had a mass-loss rate higher than $\dot{M}=10^{-3}$\,$M_{\odot}$/yr but the mass-loss slowed leading up to the time of explosion. A value of $s > 2$, however, indicates that the mass-loss rate increased with time up to $\dot{M}=10^{-3}$\,$M_{\odot}$/yr leading up to the time of explosion. Thus, the trends shown here reflect the fact that the $s<2$ CSM scenarios presented result in the supernova shock interacting with more total mass which powers a brighter light curve. As $s$ decreases, the shape of the light curve post-peak flattens as well, resulting from the sustained higher CSM densities at larger radial distances.


Decreasing $E$ also dramatically changes the light curve shape by flattening the peak and decreasing the peak luminosity. Comparing the curves at various energies to those light curves shown in Figure~\ref{fig:IIn_8GHz} already shows preliminarily that for type IIn, relatively lower energy explosions can still be responsible for bright radio emission. The ejecta density profile, however, has minimal effect on either the radio or millimeter light curves, indicating that detections of these events will not discern between different models of $n$ for the progenitor. Taking into the account all of these parameters is crucial to painting a more complete picture of the explosion and the progenitor's history

\subsection{Case Study: SN 2006jd}

\begin{figure}[t]
    \centering
    \includegraphics[width=\columnwidth]{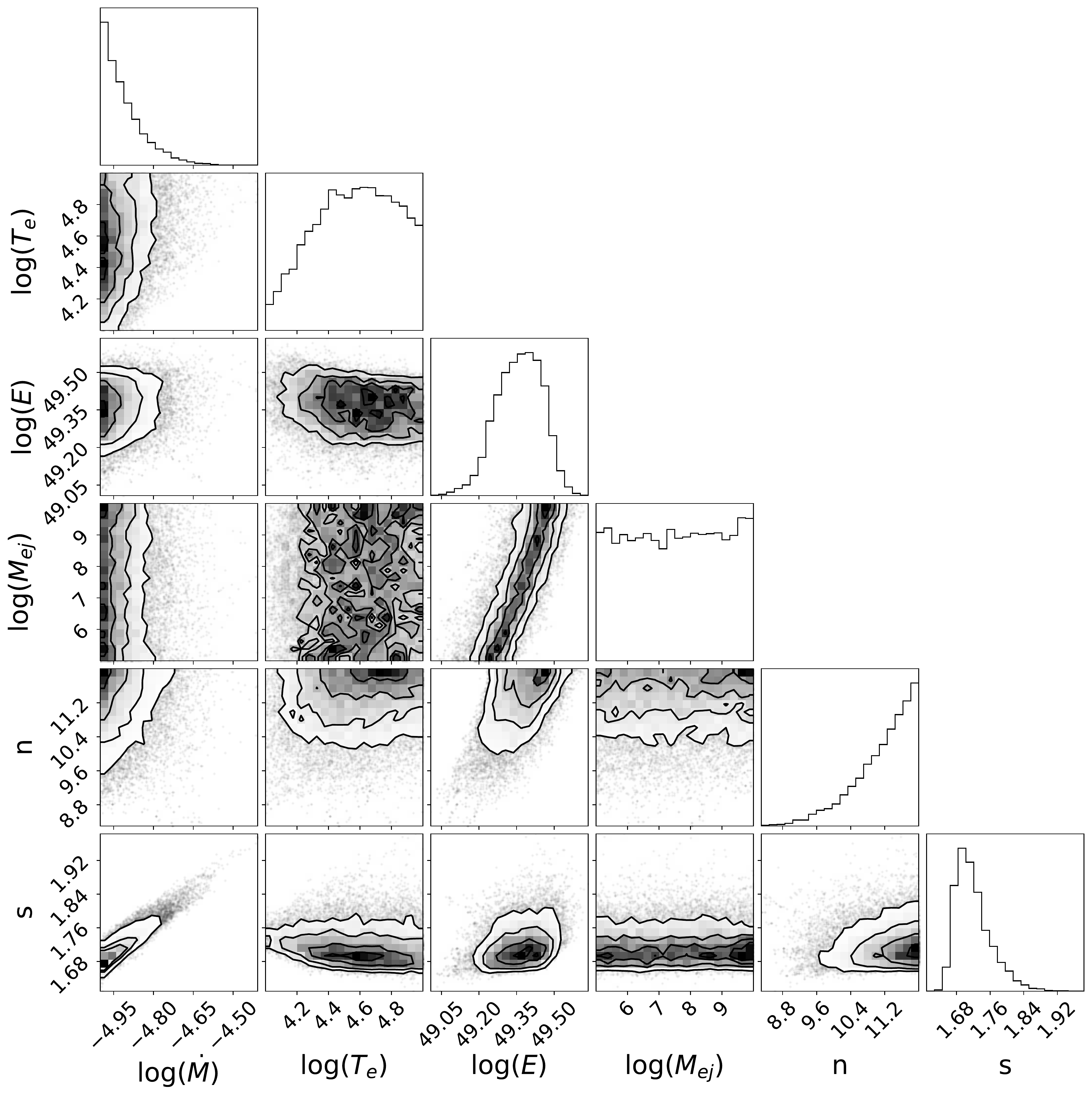}
    \caption{Corner plot showing posterior distributions of parameters used to generate a best fit model light curve for the 8.5\,GHz data of SN 2006jd. Not all of the parameters are well constrained in this analysis, indicating that late time radio data alone is not sufficient to understand these events.}
    \label{fig:06jd_corner}
\end{figure}

To illustrate an application of the modelling described in the previous section, we fit model light curves to 8.5\,GHz detections of SN 2006jd. SN 2006jd was discovered on 2006 October 12 \citep{Blondin2006}. It was originally classified as a type IIb SN and noted for its spectral similarity to SN 1993J, though it was later reclassified as a type IIn. Most type IIn supernovae have very little, if any, monitoring with radio telescopes. For 2006jd, however, \cite{Chandra2012} reports on four years of follow-up observations of the event with the Very Large Array (VLA), ranging from 400 to 2000 days post-explosion. We convert the flux densities reported in \cite{Chandra2012} to luminosities using a distance of 79\,Mpc to the event and utilize \texttt{emcee} \citep{Foreman-Mackey2013} to explore combinations of model parameters that may fit the data. For a given progenitor model, we allow parameter exploration of mass loss rate, CSM temperature, explosion energy, ejecta mass, ejecta density profile index, and CSM density profile index. We use broad tophat priors for the parameters, but we note that when applying this modelling to newly discovered events, priors should be informed by interpretation of early detections in optical and other wavelengths. The parameter posterior distributions for 2006jd assuming an RSG progenitor is shown in Figure~\ref{fig:06jd_corner}.

\begin{figure}[h]
    \centering
    \includegraphics[width=\columnwidth]{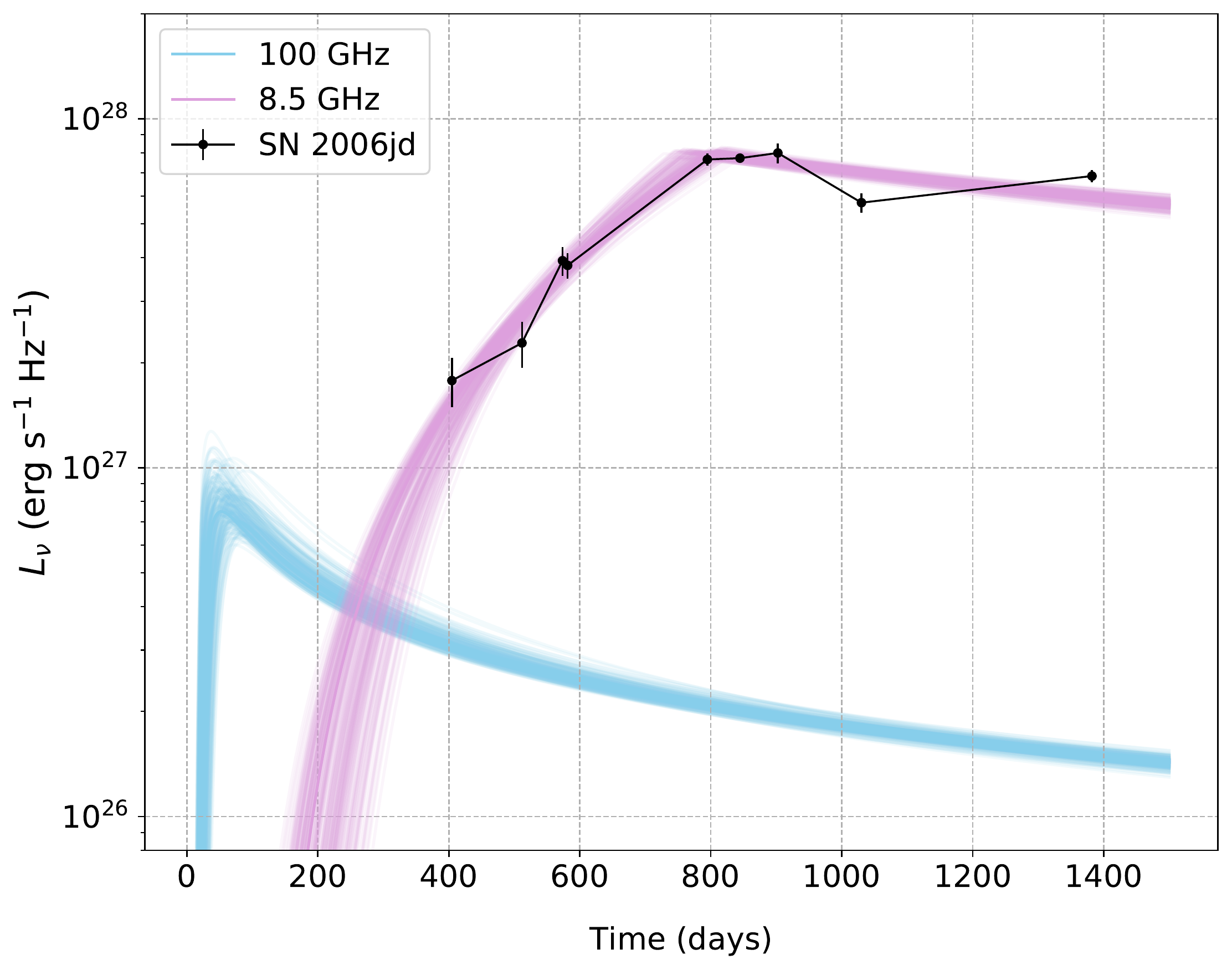}
    \caption{Light curves at 8.5\,GHz generated from 200 random samples from the MCMC fit to the SN 2006jd data. Using the same random samples, light curves are also generated at 100\,GHz. Using all 2000 samples generated in our MCMC analysis, the mean peak location of the 8.5\,GHz light curves occurs around 794 days, with a 1$\sigma$ spread of 23 days. The mean peak location of all 100\,GHz light curves occurs around 61 days with 1$\sigma = 15$ days. The potential for detecting the millimeter light curve peak demonstrated here highlights the need for sensitive and early-time follow up observations of sources such as SN 2006jd}
    \label{fig:06jd_lc}
\end{figure}

We note a few interesting characteristics of the posteriors and compare them with \cite{Chandra2012} as well as \cite{Moriya2013}. The latter work uses a similar shock evolution model as us and analytically models bolometric light curves for type IIn SNe model using a power law declining luminosity. Our results are generally consistent with both of these works. We find that a CSM density profile around $s=1.7$ is preferred for 2006jd, which matches closely with the $s=1.77$ preferred in \cite{Chandra2012} but deviates slightly from the $s=1.4$ preferred to \cite{Moriya2013}. However, the fact that all of the values are significantly below 2 indicate a shared conclusion that episodic mass-loss may be responsible for the CSM enrichment in 2006jd. We also find that the light curve models are not very sensitive to $M_{ej}$ or $n$, similar to \cite{Moriya2013} where changing these values alters the explosion energy and mass-loss rate by only around $\sim 10 \%$. 

Our results for the most likely explosion energy and mass-loss rate, however, deviate from those found in \cite{Moriya2013}. We find an explosion energy requirement of only $\sim$few $\times 10^{49}$\,ergs and a small mass-loss rate of $\dot{M} \sim 10^{-5}$\,$M_{\odot}$/yr. In contrast, the fits to the bolometric light curves of 2006jd in \cite{Moriya2013} yielded an energy requirement of $\sim 10^{52}$\,ergs and a higher mass-loss rate of $\dot{M} \sim 10^{-3}$\,$M_{\odot}$/yr. \cite{Moriya2014} estimates the mass-loss history of several type IIn SNe and indicates a slowing in the already especially low amount of mass lost from the 2006jd progenitor in the final days before explosion. Thus, fitting one fixed value of $\dot{M}$ will likely lead to an imprecise model over longer spans of time. Another source of discrepancy likely arises from the fact that these light curve models estimate the amount of energy needed to power just the synchrotron emission. As synchrotron emission traces only the fastest moving ejecta, it doesn't represent the full energy budget for the explosion. However, the true fraction of the total energy it represents is impossible to constrain. As a final note on the posteriors, the light curve model appears to weakly prefer a $T_{e} \sim 4 \times 10^4$, but we interpret this with caution as we do not expect late time data to constrain this parameter given that especially for a low mass-loss rate the shock would be in a regime that is optically thin to free-free by then.

These comparisons to \cite{Chandra2012} and \cite{Moriya2013} are meant only to confirm the validity of our posteriors, not to claim new insight on 2006jd. The aim of this section to is demonstrate the potential for observing millimeter flux from interacting supernova by inferring from previous radio detections. In Figure~\ref{fig:06jd_lc}, we show model 8.5\,GHz light curves generated using parameters from 200 random samples from the MCMC chains. For the sample parameters, we generate light curves at 100\,GHz as well. These light curves demonstrate that for parameters that generate late time emission in a fairly narrow range of possible luminosities at 8.5\,GHz, they generate early peaks at a diversity of possible luminosities for 100\,GHz. A 100\,GHz peak of $\sim 10^{27}$\,erg\,s$^{-1}$\,Hz$^{-1}$ at the distance of 2006jd corresponds to a flux density of $\sim 140$\,$\mu$Jy. A significant detection of this flux density can easily be made with facilities like the Atacama Large Millimeter/submillimeter Array (ALMA).

\section{Rates of Detection in Blind Surveys} \label{sec:rate}

Serendipitous millimeter transient detections are expected to increase in the era of wide-field CMB surveys. We explore their potential to detect type IIn supernovae in a blind search by comparing the areal densities of the supernovae with the surveys' areas and point source sensitivities. The aereal density represents the number of sources we expect to detect in the sky per square degree at any given time. Thus, it is dependent on the volumetric event rate as well as the characteristic event timescale.

To derive a volumetric rate for type IIn SNe, we rely on the Zwicky Transient Facility's (ZTF) Bright Transient Survey (BTS) which recently provided statistics for supernova demographics, including corrections for various sources of inefficiency \citep{Perley2020}. They conclude that the rate of core-collapse supernovae is $1.01 \times 10^5\ $\,Gpc$^{-3}$\,yr$^{-1}$ and that 10.2\% of those are classified as type IIn. Thus, we adopt a type IIn supernova rate of $1.03 \times 10^4\ $\,Gpc$^{-3}$\,yr$^{-1}$. We consider supernovae with RSG and BSG progenitors and three different mass-loss rates, assuming fixed parameters of $T_e = 10^4$\,K, $E=10^{51}$\,ergs, $M_{ej}=5$\,$M_{\odot}$, n=10, and s=2. 

\begin{figure}[t]
    \centering
    \includegraphics[width=\columnwidth]{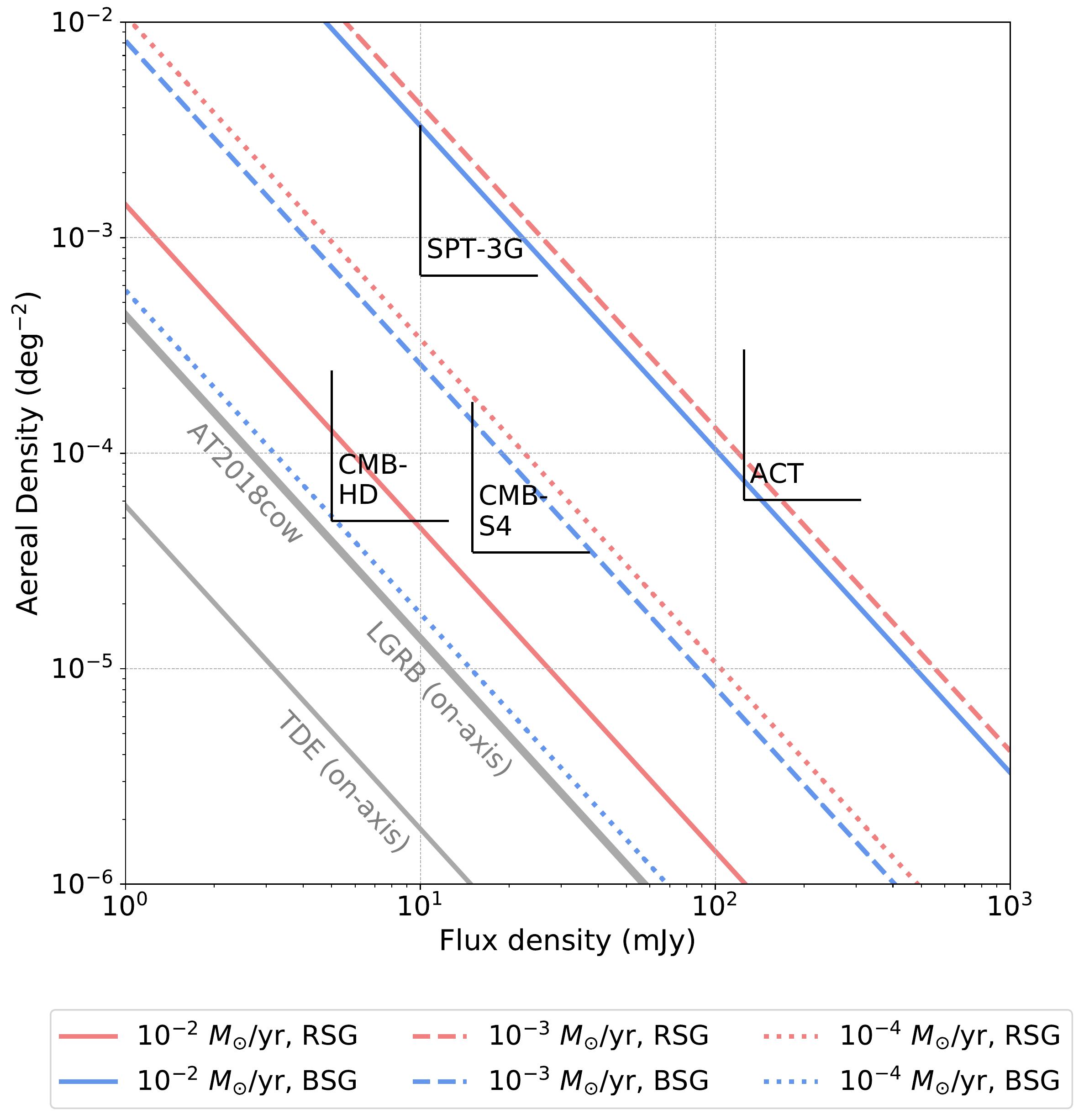}
    \caption{Areal densities of interacting SNe as a function of their flux densities for various progenitor models. Three mass loss rates are depicted, $\dot{M}=10^{-2}$\,$M_{\odot}$/yr, $\dot{M}=10^{-3}$\,$M_{\odot}$/yr, and $\dot{M}=10^{-4}$\,$M_{\odot}$/yr. All models have values of $T_e = 10^4$\,K, $E=10^{51}$\,ergs, $M_{ej}=5$\,$M_{\odot}$, n=10, and s=2. Also shown are limiting aereal densities for the 5$\sigma$ detection thresholds for four different CMB surveys. For comparison, aereal densities of other millimeter-bright astrophysical events are also included. The line for AT2018cow is generated using values given in \cite{ho2021}, while the lines for on-axis long gamma-ray bursts (LGRB) and on-axis tidal disruption events (TDE) use values given in \cite{Eftekhari2021}.}
    \label{fig:rates}
\end{figure}

For each model, we set the characteristic timescale of the light curve by computing the amount of time between the luminosity reaching half the peak luminosity. We can compute the flux densities we expect to observe for a range of distances. The distances we consider are within a few hundred Mpc, chosen to roughly correspond to the flux density sensitivity limits of relevant millimeter surveys. Using the IIn SNe event rate given above, we can compute the number of supernovae per year that exhibit a half-peak brightness above a given flux density. Multiplying  by the light curve timescale and dividing by the total number of degrees in the full sky then yields aereal density. Figure~\ref{fig:rates} shows these lines of supernova areal density as a function of half-peak flux density for the various models.

In Figure~\ref{fig:rates}, we also show the limiting areal densities and flux densities of four CMB surveys of interest. Surveys are sensitive to sources with areal densities greater than 1/FOV\,deg$^{-2}$ at a $5\sigma$ detection. The SPT observes a 1500\,deg$^2$ area of sky down to $\sigma=2$\,mJy in their 95\,GHz band (S. Gunns, 2021, personal communication) while the ACT observes 40\% of the sky every week down to a $\sigma=25$\,mJy in their 90\,GHz band (K. Huffenberger and S. Naess, 2021, personal communication). The SPT is within the threshold of being to observe supernovae with progenitors undergoing amongst the most extreme mass-loss cases we consider in this work, while the ACT is unlikely to observe anything. This is consistent with the rate and types of extragalactic transients these surveys have already found. CMB-S4 is a next generation CMB survey with stations in both Chile and the South Pole. It is expected to conduct deep observations of 70\% of the sky, covering frequencies between 30-280\,GHz \citep{Abazajian2019}. The experiment also expects to deliver observations of transient sources using a difference imaging pipeline. At 95\,GHz, CMB-S4 will achieve $\sigma=3$\,mJy in a one week stack. Looking further to the future, CMB-HD is a new experiment proposed for the Astro2020 Decadal Survey. Projected to have more frequency bands and observe deeper, CMB-HD expects to observe 50\% of the sky and achieve $\sigma=1$\,mJy for detecting transients. We expect that these two next generation CMB experiments to have the potential to detect bright and local type IIn supernovae, as they lie within the requirements of sensitivity and areal density to observe a broader range of progenitor mass-loss rates. 

We caution that the supernova areal densities we present are likely optimistic. Several progenitor scenarios may contribute to the population of type IIn supernovae that have been observed to date, and we do not believe that any of these lines of areal density are representative of the true sample of events. Additionally, we have assumed that $\epsilon_e = \epsilon_b = 1/3$. Lower energy fractions could yield significantly lower light curve luminosities, decreasing the rates of detection we predict for these surveys. We also assumes homogenous CSM structure when in reality, CSM interaction is often complicated by inhomogenous structure and asymmetry and may require more sophisticated modelling to properly describe it. Given this, blind surveys should be able to detect a sample of the most nearby and most extreme type IIn SNe. Follow up millimeter observations of optically discovered SNe would supplement these detections, shedding light on the range of possible progenitors and mass-loss histories that produce these types of events.

\section{Conclusion} \label{sec:conclusion}

The main goal of this work was to demonstrate the utility of early millimeter observations of supernovae interacting with dense circumstellar environments. A variety of progenitor models have been proposed for these events, but discerning between different theories requires knowledge of the progenitor's pre-explosion behavior. Probing the CSM into which a supernova explodes can give us insight into the final days of stellar evolution for massive stars and better constrain the viability of different progenitor models.

Here, we apply a synchrotron emission model that takes into account slow versus fast cooling regimes and free-free absorption to produce millimeter and radio light curves of interacting SNe. The light curves are generated by using the prescription of shock radius and velocity evolution outlined in \cite{Chevalier1982} and assuming that a constant fraction of the total shock energy goes into electrons and magnetic fields. The model light curves of the first 1500 days post-explosion reveal that millimeter emission peaks significantly sooner than lower frequency radio emission, making it an important probe of the CSM density close to the surface of the progenitor. The light curves are most sensitive to changes in explosion energy and CSM density profile but are fairly insensitive to changes in the ejecta density profile. 

For a practical application, we utilize \texttt{emcee} to fit parameters of synchrotron-powered light curve models to an 8\,GHz light curve of type IIn SNe 2006jd. We find that the light curve was most likely generated by a low energy explosion and a relatively low amount of mass-loss enriching the CSM but with a profile indicative of non-steady mass-loss. The model fitting was not very sensitive to the CSM temperature or the total ejecta mass. Millimeter light curves generated by random samples of parameters from the analysis show the possibility of an early and luminous peak with potential for detection by sensitive instruments such as ALMA. 

In the upcoming era of wide-field CMB surveys, we believe there is potential for blind detections in the millimeter band of especially luminous and nearby interacting SNe. Combining these with millimeter followup observations of SNe discovered in optical surveys, a representative sample of interacting SNe can be built. Given millimeter light curves, fitting to the models described here presents a method to solve for first-order estimates of the viable progenitor and CSM parameters that describe the emission. Interpreting their millimeter light curves in conjunction with multi-wavelength observations will be key in broadening our understanding of the evolution and mass-loss histories of massive stars that lead to these events.

The code used to generate the figures shown in this paper can be found at \url{https://github.com/nitikayad96/mmbrightsupernovae.}

\section{Acknowledgements}

We thank Sterl Phinney and Abigail Polin for useful discussions in formulating the model for this work. We also thanks Sam Guns for discussions regarding the SPT-3G as well as Sigurd Naess and Kevin Huffenberger for discussions regarding the ACT. This research has made use of NASA’s Astrophysics Data System Bibliographic Services. 

\bibliography{mmSN}{}
\bibliographystyle{aasjournal}

\end{document}